\documentclass[fleqn,12pt,twoside]{article}
\usepackage{espcrc1}

\usepackage{graphicx}

\def\Journal#1#2#3#4{{#1} {\bf #2}, #3 (#4)}

\def\NPA{{\em Nucl.~Phys.} A}
\def\PLB{{\em Phys.~Lett.}  B}
\def\PRL{\em Phys.~Rev.~Lett.~}

\def\PRC{{\em Phys.~Rev.} C}
\def\ZPC{{\em Z.~Phys.} A}
\def\ZPC{{\em Z.~Phys.} C}

\def\JP{{\em J.~Phys.} G}

\newcommand{\AmS}{{\protect\the\textfont2
  A\kern-.1667em\lower.5ex\hbox{M}\kern-.125emS}}

\title{Collective effects in microscopic transport models}

\author{Carsten Greiner\address{Institut f\"ur Theoretische Physik I,
Universit\"at Giessen, D-35392 Giessen, Germany}}
       
\begin{document}

\maketitle

\begin{abstract}
We give a reminder on the major inputs
of microscopic hadronic transport models and on
the physics aims when describing various aspects
of relativistic heavy ion collisions at SPS energies.
We then first stress that the situation of particle ratios being
reproduced by a statistical description does not necessarily mean
a clear hint for the existence of a fully isotropic momentum distribution
at hadrochemical freeze-out.
Second, a short discussion on the status of strangeness production
is given. Third
we demonstrate the importance of a new collective mechanism for producing
(strange) antibaryons within a hadronic description, which
guarantees
sufficiently fast chemical equilibration.
\end{abstract}

\section{PHILOSOPHY OF HADRONIC TRANSPORT MODELS}

Microscopic hadronic transport models have guided
in detail the understanding and interpretation
of experimental results of the heavy ion
programs at Bevalac, SIS, AGS and SPS energies.
For these reasons one can say that this has been the {\em
golden age} of those models.
Competing approaches are the (R)QMD model(s),
the URQMD model, the HSD model, the
RBUU model(s) and the QGSM model, all including
the subsequent microscopic interactions of the hadrons
as the common collective feature underlying the dynamics.

These models do provide a complete space-time picture
(including momentum space) of a
violent heavy ion reaction.
They can describe the buildt up of collective
effects like flow observables, the
production of all kinds of particles, like strangeness and
the production of charmonium states, and
also electromagnetic probes.
The underlying dynamics
is described by a phenomenological reaction network
of various hadronic Boltzmann processes.
As the basic input are vacuum cross sections
for the various elastic and inelastic hadronic reaction channels,
semiclassical transport approaches do have
a profound and solid foundation: relativistic kinetic theory
(being, in principle, superior to any thermal or hydrodynamical description)
and known vacuum physics.

However, caution enters into the description
when cross sections are asked which can not be measured experimentally.
Besides the dynamics of all kinds of hadronic resonances
and their implementations this is, in particular,
true for the modelling of the highly energetic inelastic binary hadronic
reactions (by eg string excitation and fragmentation) and their collective
incorporation into space-time dynamics.
This is the situation where the
transport description has a character
of a model, as the basic input
for the very initial stage of the reaction
relies on collective, nonperturbative and non-equilibrium QCD, which
can only be mimiced by phenomenological concepts.
One has to describe a regime which is dominated by
soft processes, and which might well also be the situation
encountered still at RHIC energies.

In this respect it is clear that transport approaches
can not really implement any (pre-) quark gluon plasma (QGP) state of
deconfined matter. One can turn
this major deficiency into a positive perspective:
One can look for observables to see if
they show a significant dependence on physics
incorporated by the description of the very initial stage of the
reaction: Is there any new physics necessary?!
In this respect with the transport description at hand
one can also implement possible in-medium modifications
of particle properties during the course of the reaction
and hence test for experimental obervation possibilities.

Momentum equilibration (`thermalization') and
chemical equilibration of the various particle species
can thoroughly be tested \cite{Bravina,BCG00}.
On the other hand, by the success of applying
`thermal' or statistical models (eg \cite{BMS96}) to describe
the final hadronic yield, it became quite popular to belief
that the applicability of such
approaches also proves for the existence
of a state of almost complete thermal equilibrium.
We will stress in section \ref{sec2}
that this does not necessarily mean
a clear hint for the existence of a fully isotropic momentum distribution
at hadrochemical freeze-out. Therefore the studies
of equilibration by hadronic transport models are
necessary for a detailed understanding of the reaction
dynamics.
Second, we will give a brief discussion
in section \ref{sec3} on the status of
strangeness production within the models, also
refering to an ongoing analysis. Strangeness production
had been raised as one of the potential QGP signatures,
and here especially rare antihyperons as the important messenger
of the initial phase.
Refering to the later, section \ref{sec4} is devoted
to a new collective mechanism for producing
(strange) antibaryons within a hadronic description \cite{GL00}.
So far, the description of the (strange) antibaryon
production within standard hadronic transport schemes
in comparison to data had faced notorious difficulties.
As the new idea has been presented already at the last Quark Matter
conference, novel numerical calculations for a dynamical setup are
reported.

\section{STATISTICAL MODEL WITH ANISOTROPIC MOMENTUM
\\
DISTRIBUTIONS}
\label{sec2}

Besides the microscopic transport approaches
global, statistical analyses as
complimentary theoretical descriptions
of the final hadronic yield have been explored
in detail over the last
years (see eg \cite{BMS96}).
The `thermal' model description of a noninteracting hadronic
resonance gas works astonishingly well to reproduce
the various hadronic abundancies by fixing only two
intensive parameters, the temperature $T$ and baryochemical potential
$\mu_B$.
From the many analyses one is generally tempted to conclude
that at the point where the picture of
chemical composition (or `freeze-out') does apply,
a complete thermal state has been established with
locally having isotropic distributions in momentum space.
On the other hand transport models do not really support the
idea that a complete local momentum equilibration
can be achieved even at the later stages of the reaction \cite{Bravina,BCG00}.

\begin{figure}[t]
\begin{minipage}[t]{78mm}
\centering
\includegraphics[width=49mm]{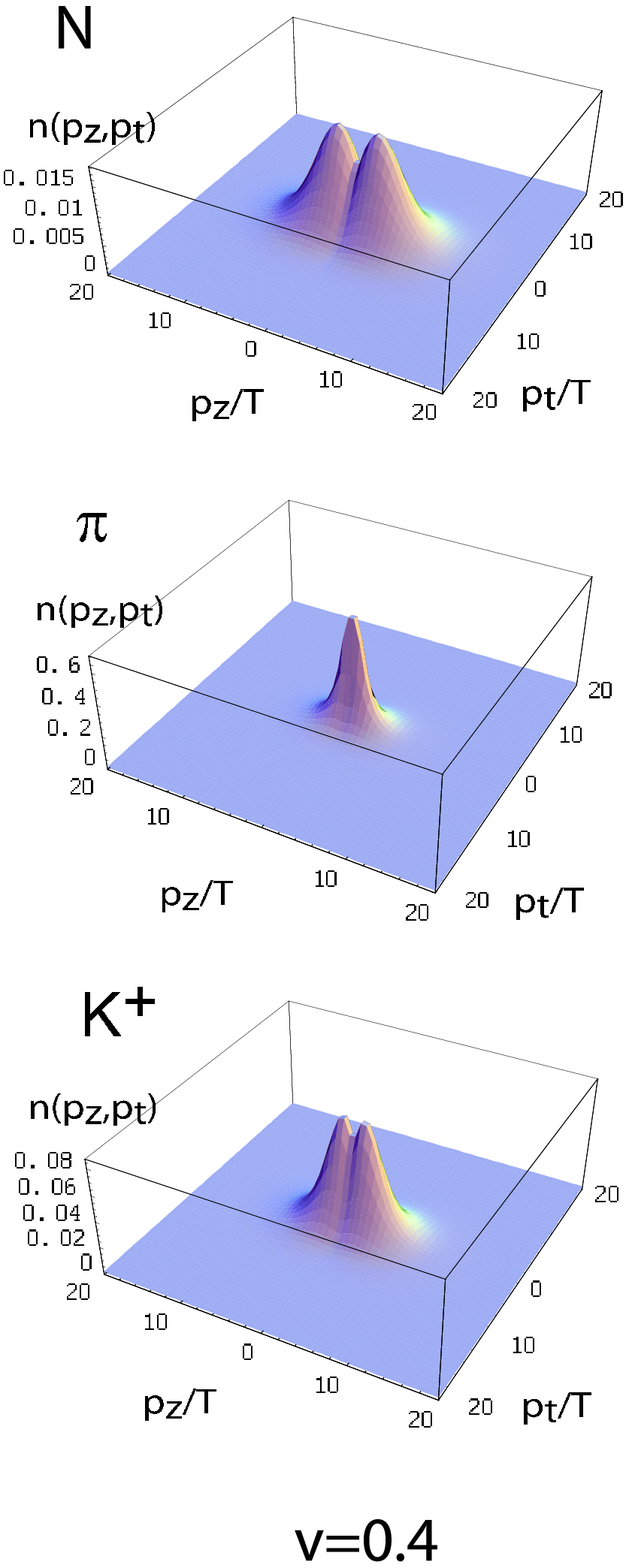}
\centering
\caption{Generalized statistical phase space distribution
for $v=0.4$. Temperature parameter and chemical potentials
are taken as those extracted for $v=0$ being typical for Pb+Pb collisions.}
\label{fig_aniso}
\end{minipage}
\hspace{\fill}
\begin{minipage}[t]{78mm}
\centering
\includegraphics[width=80mm]{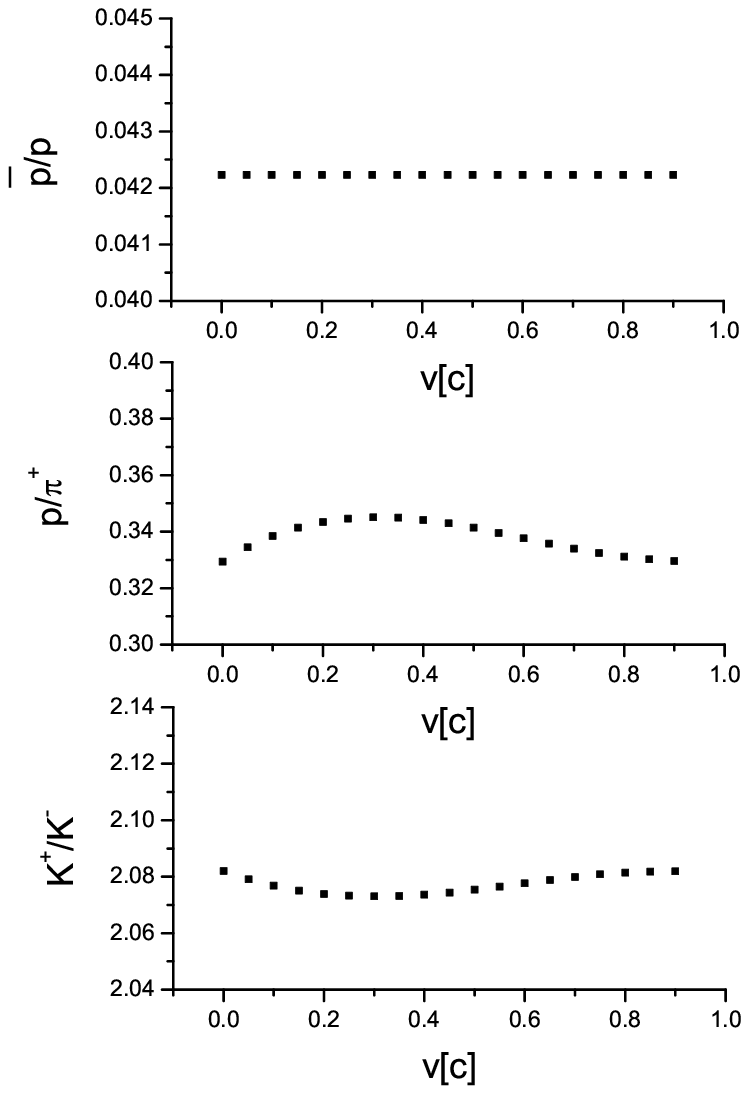}
\centering
\caption{(Weak) Dependence of typical particle ratios
varying the velocity parameter.
Temperature parameter and chemical potentials
are taken as those extracted for $v=0$.}
\label{fig_rat}
\end{minipage}
\end{figure}

The resolution to this apparent conflict is to relax the
constraint of assuming fully isotropic, statistical distributions
for comparison to hadron abundancies, i.e. of assuming complete
thermal equilibrium \cite{SG02}.
At the onset of the heavy ion reaction the local momentum
distribution of the
hadronic particles is far from being isotropic. Such a mismatch
should relax due to collisions among the particles, though
the system might never actually become locally completely
isotropic  \cite{Bravina,BCG00}. To allow for momentum anisotropy
within a statistical concept, one has to start from the statistical density
operator
\begin{equation}
    \rho_{GK} \, = \, \frac{1}{\Omega}\exp(-\beta(H-\mu N-v O)) \, \, ,
\end{equation}
where $O$ defines the adequate one-particle operator for momentum anisotropy
and $v$ is the conjugated intensive quantity \cite{Neise}.
For stationarity, the 'von Neumann-equation'
$\frac{d \rho}{dt}=\{H,\rho\}=0 $ is fullfilled, if $O$ commutes
with the (effective) Hamiltonian.
If one defines the momentum anisotropy $O$ as the
difference between the mean value of the momenta with a positive
and negative longitudinal ($z$) component, respectively,
i.e.
$ \hat{O}(\hat{p})=\sum_p |p_z|\hat{a}_p^+\hat{a}_p $,
the requirement for stationarity is fulfilled when employing
effective Hamiltonians of mean-field type or a simple quasi-free
Hamiltonian of a hadron resonace gas. The incorporation of binary collisions,
of course, will destroy stationarity.
Assuming a noninteracting hadron
(Hagedorn-like) resonance gas,
as typically employed in all analyses, and
minimizing the grand canonical potential $\Omega $
leads to the generalized momentum distribution
\begin{equation}
    \rho_i \, = \, \frac{g_i}{(2 \pi)^3}\int d^3p \frac{1}{\exp(\beta
    \gamma (E_i-v|p_z|)-\beta \mu_{i})\pm 1} \, , \label{formel}
\end{equation}
with $\gamma = (1-v^2)^{-1/2}$.
The intensive parameter $v$ can be interpreted as the
relative velocity between projectile and target.
Fig.~\ref{fig_aniso} illustrates examples
for the case of nucleons, pions
and kaons with the intensive parameter
$v=0.4$ c.

In Fig.~\ref{fig_rat} now the actual (non-)dependence
of exemplaric particle ratios on varying the velocity parameter is depicted.
The temperature parameter and chemical potentials
are taken as those extracted for $v=0$ (i.e. being
compatibel with those of \cite{BMS96}).
The largest deviation, about 6\%, one finds
for the $p/\pi^+$ ratio (for the situation of a Pb-Pb collisions at SPS).
Hence the quality of the fits is not really changed
when employing any of these generalized distributions.
The same situation of encountering only a very weak dependence
is met for an analysis concerning AGS or RHIC data \cite{SG02}.

The presented results lead to the conclusion, that the succesful description
of particle ratios within a standard statistical model does not give
a proof
for the stringent existence of
local isotropic momentum distribution at
hadrochemical freeze-out, but is merely an assumption.
The achievement of momentum equilibration
can only be judged by considering hadronic transport models
in comparison to data.

\section{REMARKS ON STRANGENESS PRODUCTION}
\label{sec3}

\begin{figure}[htb]
~\vskip +2mm
\begin{minipage}[t]{67mm}
\includegraphics[width=67mm]{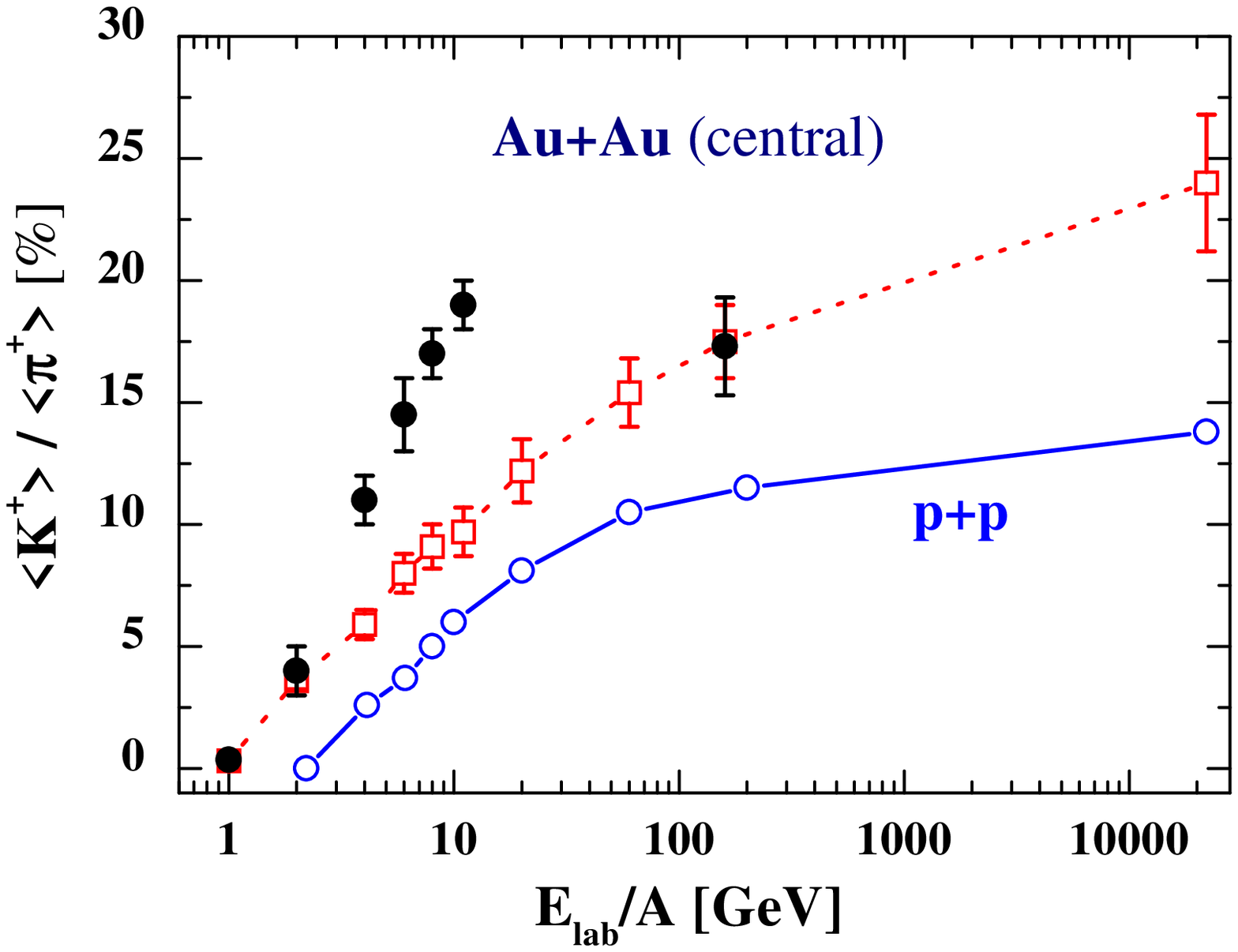}
\caption{Excitation function for $K^+/\pi^+$
around midrapidity
for central Au+Au reactions (open squares) from SIS to RHIC energies
obtained within the HSD model
in comparison to experimental
data and elementary p+p collisions (open circles).}
\label{fig_jo}
\end{minipage}
\hspace{\fill}
\begin{minipage}[t]{78mm}
\includegraphics[width=75mm]{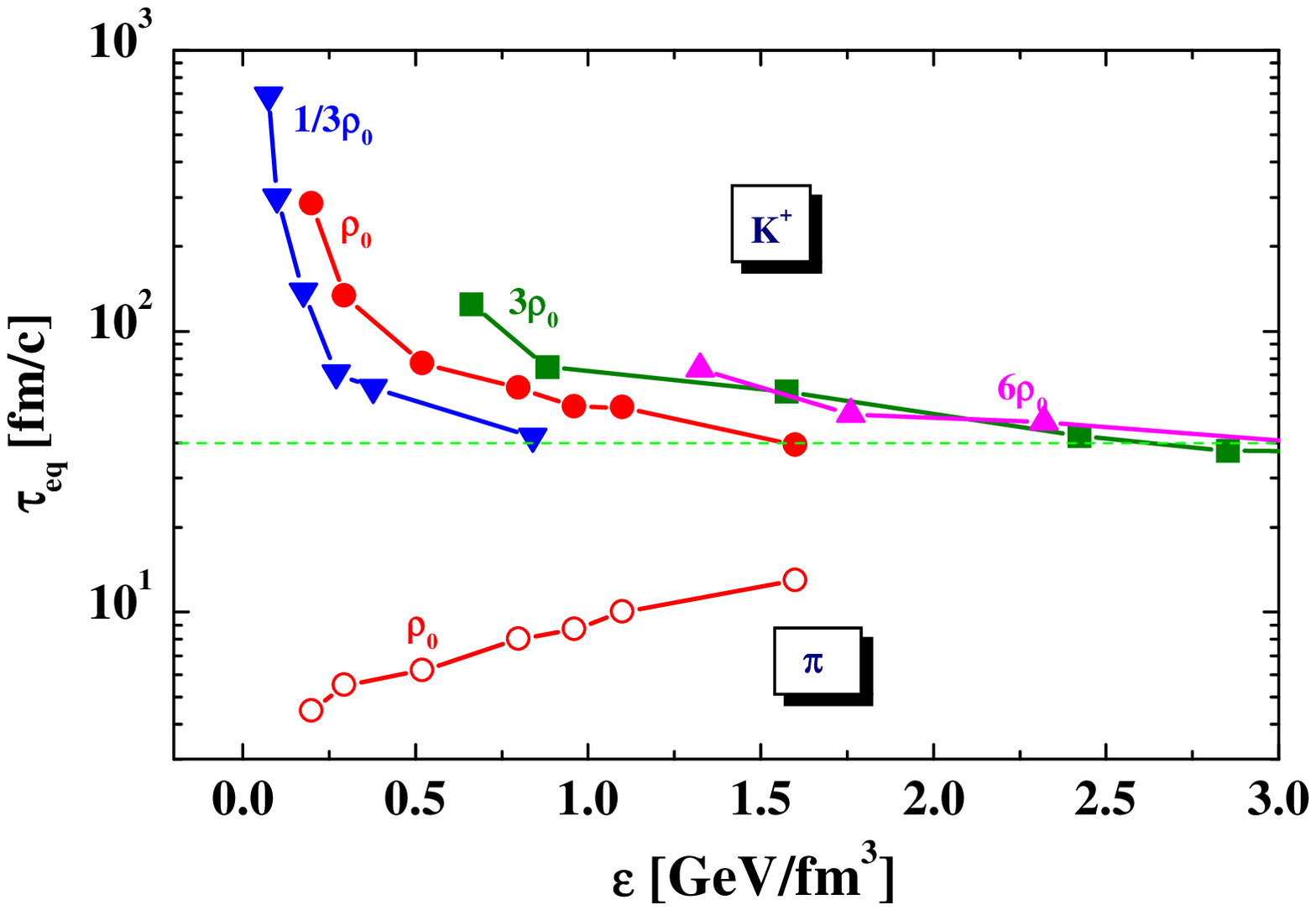}
\caption{Chemical equilibration time versus energy density
for $\pi $ and $K^+$ mesons at various different baryon densities
\protect\cite{BCG00}.}
\label{fig_el}
\end{minipage}
~\vskip -4mm
\end{figure}

There have been a couple of detailed studies to
understand strangeness production for AGS and SPS energies
within a specified microscopic model
(for reviews see \cite{So98,Bass,CG01}).
To be specific, in \cite{Ge98} the properties of $K^+ $, $K^-$ and
$\Lambda $ particles in nuclear reactions from SIS to CERN-SPS energies
have been investigated within the HSD model.
Here also available p+A data has been confronted.
The outcome for the excitation function of the $K^+$-mesons,
relative to the $\pi^+$-yield
is summarized in  fig.~\ref{fig_jo}.
After the primary string fragmentation
of intrinsic p-p--collisions the hadronic fireball starts with a
$K^+/\pi^+$ ratio still far below chemical equilibrium with $\approx 6 - 8 \% $
at AGS to SPS energies before the hadronic rescattering starts.
Secondary
(meson-baryon) and ternary (meson-meson)
induced string-like interactions do then contribute significantly
to additional strange particle production, particular for reactions
at SPS energies.
Via these channels about the same number of strange and anti-strange quarks
is produced as in the primary p+p collisions. This then can explain the
relative enhancement compared to p+p
(see fig. \ref{fig_jo}).
It is valid to say that the so called strangeness enhancement
at SPS energies does not require deconfinement.

The major amount
of produced strangeness (kaons, antikaons and $\Lambda $s) at SPS-energies
can be understood in terms of early and still energetic,
non-equilibrium hadronic interactions.
Afterwards, more or less
no further strangeness is being produced in any of the
microscopic hadronic transport simulations.
Putting it differently, when the
momenta of the nucleons have sufficiently degraded and the system
has to some extent thermalized, the timescale for production of
strange particles via the considered kinetic reactions becomes much too
large.
Fig.~\ref{fig_el} summarizes this statement:
The chemical saturation of the kaons has been investigated
microcanonically for a static box
and is found to be larger than 40 fm/c
for all equilibrium energy densities up to 2-3 GeV/fm$^3$
and thus exceeds considerably the lifetime
of the fireball \cite{BCG00}.

Focussing on lower bombarding energies, a smooth and continously
smaller $K^+/\pi^+$ is seen in the simulation (compare Fig.~{\ref{fig_jo}),
which clearly is not in line with the behaviour seen at AGS
energies or at lower SPS energies!
Strangeness production is underestimated by $\approx $ 30\%
compared to explicit kaon data \cite{Ge98} wheras the pion population
is in addition slightly overestimated.
The outcome of three microscopic transport simulations
has been reported in \cite{Bass}.
URQMD calculations do find agreement for the lower AGS energies, but
otherwise appreciably do underestimate the ratio.
However, again when looking on the pion yield and the kaon yield seperately,
one realizes that the difference to the experimental
rapidity distributions
is still sizeable, but not as dramatic \cite{Weber}.
Here, at SPS energies and for highest AGS energies
the pions are slightly overestimated, whereas the kaons are
slightly underestimated, the two and opposite tendencies then
are enlarged when showing only the ratio.

One conclusion thus is that showing the $K/\pi $-ratio can
distort the results, which actually should be better compared seperately.
The second and important conclusion is that one really has to understand the
physics reasons for the minor, but
still significant difference in the results
(like pion and kaon distributions)
among the various
transport approaches, as the underlying principle philosophy is the same.
It could be, on the one hand,
that the primary and violent stage of the reaction
is modelled differently, which then might lead to noticeable differences
in some observables.
As for the AGS energies and the lower SPS energies the baryon densities
achieved are the highest, especially then the baryonic resonances
could be of crucial importance and thus
could also be a possible source for the reported
differences. Hence, on the other hand,
it might be that also the treatment and thus
the understanding of the later hadronic stage of the reaction
is of significant relevance.
A detailed comparison is presently carried out by Bratkovskaya
and Weber \cite{Weber}.

\section{(STRANGE) ANTIBARYON PRODUCTION}
\label{sec4}

In particular because of high production thresholds in
binary hadronic reaction channels
antihyperons had been advocated as the
appropriate QGP messengers.
Indeed, a satisfactory picture
of nearly chemically saturated populations of
antihyperons has been demonstrated over the
last years with the Pb+Pb experiments NA49 and WA97 at CERN-SPS
\cite{BMS96}.
On the other hand the theoretical description of the antibaryon
production within hadronic transport schemes
in comparison also to earlier data for light systems
faced some severe difficulties.
Phenomenological motivated attempts \cite{So95}
to explain a more abundant production of antihyperons had been
proposed like the appearance of color ropes,
the fusion of strings,
the percolation of strings, or the formation of high-dense
hadronic clusters.
Their purpose is mainly
to create (much) more antibaryon in the very early
intial stage of the reaction (compared to simple
rescaled p+p collisions).
Still, in most of the transport calculations
a dramatic role of subsequent antibaryon annihilation
is observed, which, in return, has to be more than counterbalanced
by the initial production.
The situation seems even more paradox with respect to the fact that
the chemical description by thermal models works indeed
astonishingly well
for the antihyperons and being, of course, completely nondependent on
the (large) magnitude of the annihilation cross section.
For all of this reasons the
theoretical and dynamical understanding of the
production of (strange) antibaryons has remained a delicate and challenging
task \cite{So98}.

\begin{figure}[htb]
\begin{minipage}[t]{78mm}
\includegraphics[width=92mm]{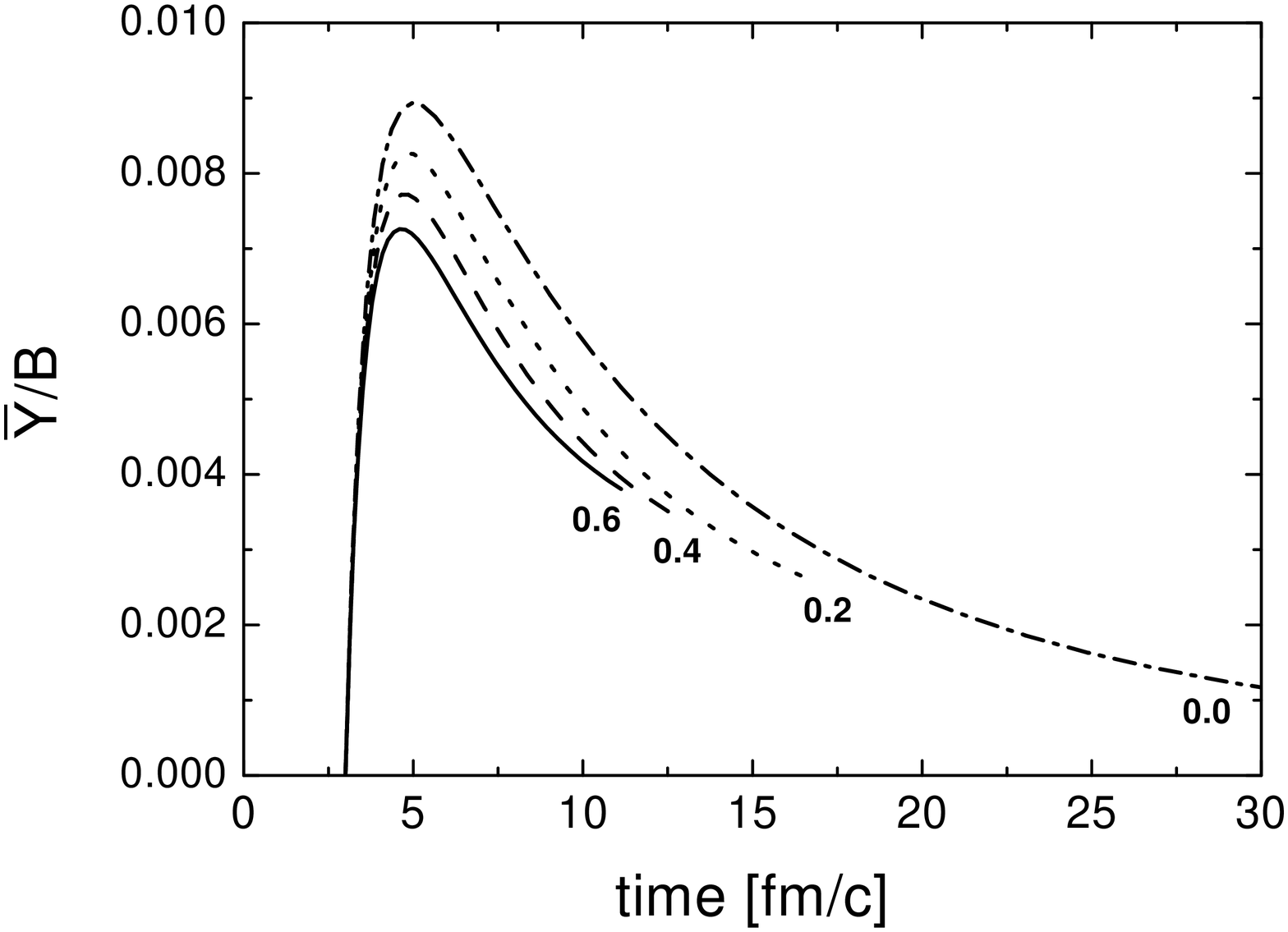}
~\vskip -9mm
\caption{
        The anti-$\Lambda $ to baryon number ratio
$N_{\bar{ \Lambda }}/N_B \, (t)$
as a function of time for various velocity
parameters $v_{lin}$ for the transverse expansion.
The entropy per baryon is taken as $S/A=30$, $t_0=3 $ fm/c and
$T_0=190 $ MeV.
}
\label{fig_vlin}
\end{minipage}
\hspace{\fill}
\begin{minipage}[t]{78mm}
\includegraphics[width=92mm]{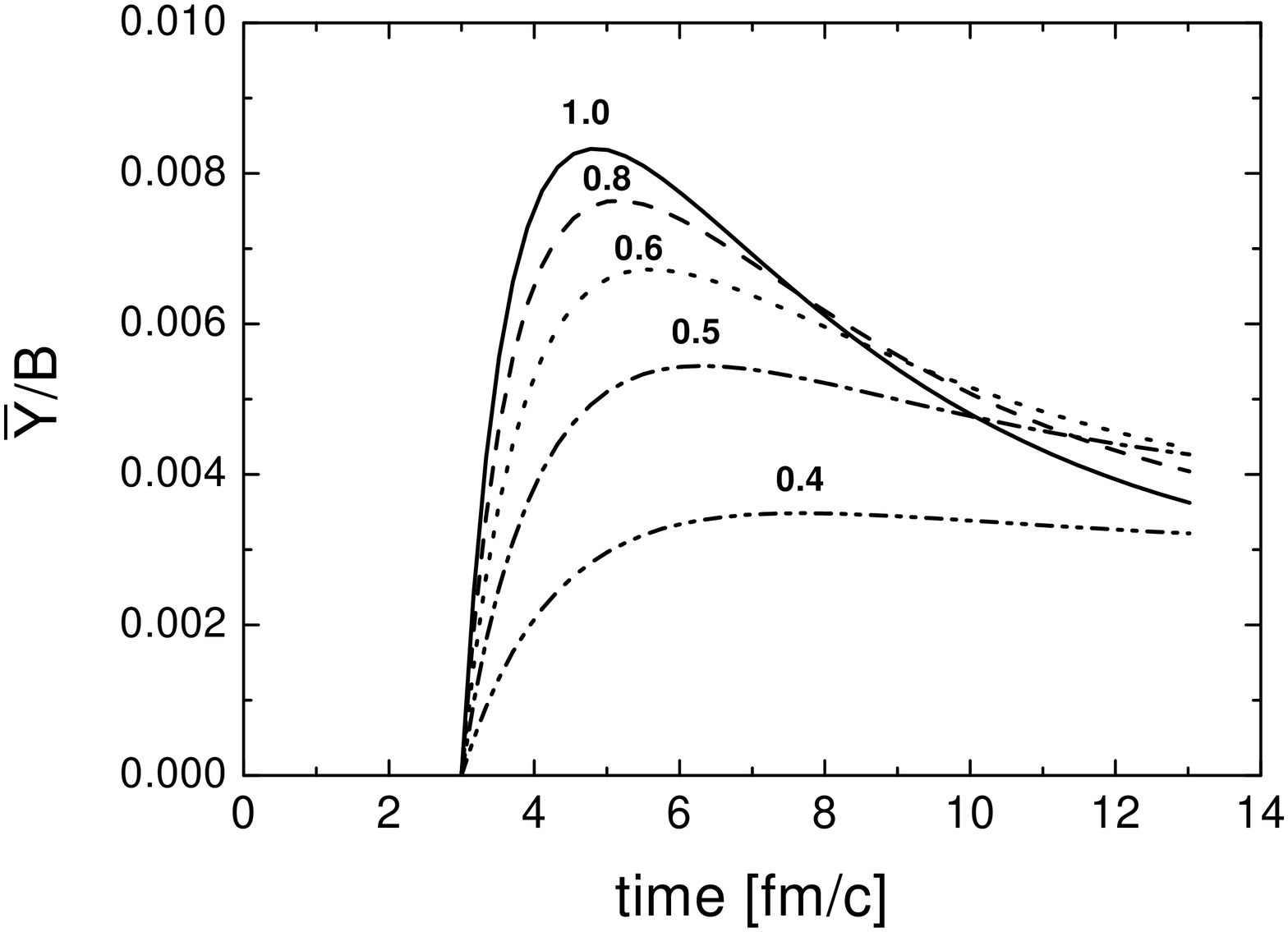}
~\vskip -9mm
\caption{
        The anti-$\Lambda $ to baryon number ratio
$N_{\bar{ \Lambda }}/N_B \, (t)$
as a function of time for various
implemented annihiation cross section $\sigma_{eff} \equiv \lambda \,
\sigma_0 $.
The entropy per baryon is taken as $S/A=30$, $t_0=3 $ fm/c and
$T_0=190 $ MeV.
}
\label{fig_lam}
\end{minipage}
\end{figure}

Amazingly, it was noticed only recently that
a correct incorporation of the baryonic annihilation
channels had actually not been done consistently in the
simulations \cite{RS00,GL00}.
A sufficiently fast redistributions of strange and light quarks
into (strange) baryon-antibaryon pairs should be achieved
by multi-mesonic fusion-type reactions of the type
\begin{equation}
\label{mesfuse}
n_1\pi + n_2 K \, \leftrightarrow \, \bar{Y}+p
\end{equation}
occuring in a moderately dense hadronic system \cite{GL00}.
Naively one thinks that the probability
of multiple pions and kaons to come close in space
is very low and therefore irrelevant.
However, (at least) these special kind of multi-hadronic
reactions have to be present because of the fundamental principle of detailed
balance.

In the following we present
rate calculations for a dynamical setup of an expanding
(and nearly equilibrated) hadronic resonance gas.
Ideally the rate equations do correspond to a coarse grained,
effective description of microscopic hadronic transport.
We will turn to brief discussion of how to implement
such reactions within a microscopic, local transport
at the end of this section.
For the expansion either a linear profile
$R(t)=R_0 + v_{lin}(t-t_0)$
or an accelerating profile
$R(t)=
R_0 + v_0(t-t_0) + 0.5 a_0 (t-t_0)^2 $
is employed for the effective volume expansion
$V(t) \sim t* R^2(t)$.
The evolution of the hadronic gas
is taken to be an isentropic expansion
being specified via the entropy per baryon ratio $S/A$.
Temperature and baryon density are then evaluated as a function of
time, once an initial temperature and time has been chosen
(for more details see the
last reference of \cite{GL00}).

In Fig.~\ref{fig_vlin} the number of $\bar{\Lambda }$s
(normalized to the conserved net baryon number)
as a function of time is depicted.
As a minimal assumption the initial abundancy of antihyperons
is set to zero in all of the following figures (and actually
will not depend on the initial value (!)).
The entropy per baryon is chosen as $S/A=30$ being
a typical value to global (`$4\pi $') SPS results.
The parameter $v_{lin}$ is varied to simulate slow or fast
expansion of the late hadronic fireball.
The general characteristics is that first the antihyperons
are dramatically being populated, and then in the very late
expansion some more are still being annihilated, depending on
how fast the expansion goes. A rapid expansion gives a higher yield,
which can increase the final yield by a factor of 2 to 3.
Still, the typical expansion behaviour obtained
from simulations or extracted from the analysis of transverse
momentum slopes of pions and protons is that
at the late stages the transverse expansion velocity
should be about $0.5 $ c.

In Fig.~\ref{fig_lam} the number of $\bar{\Lambda }$s
as a function of time is depicted, where now
the annihilation cross section employed
is varied by a constant factor, i.e.
$\sigma_{eff} \equiv \lambda \, \sigma_0$ \cite{GL00}.
For the volume expansion an appropriate transverse radius
accelerating in time is now employed.
The results are rather robust against a variation by a factor of 2
in the cross section. Typically (for $\lambda =1$)
about more than 5 times
in number of antihyperons are created during the evolution compared
to the final number freezing out, thus reflecting the
fast back (`annihilation') and forth (`creation')
processes at work dictated by detailed balance.
The mass law action enforces the hyperons to
buildt up and then maintain chemical equilibrium for a certain time,
until at moderately low particle densities
the multi-mesonic fusion processes are getting ineffective
due to the rapid dilution.

\begin{figure}[htb]
\centering
   \includegraphics[height=9cm]{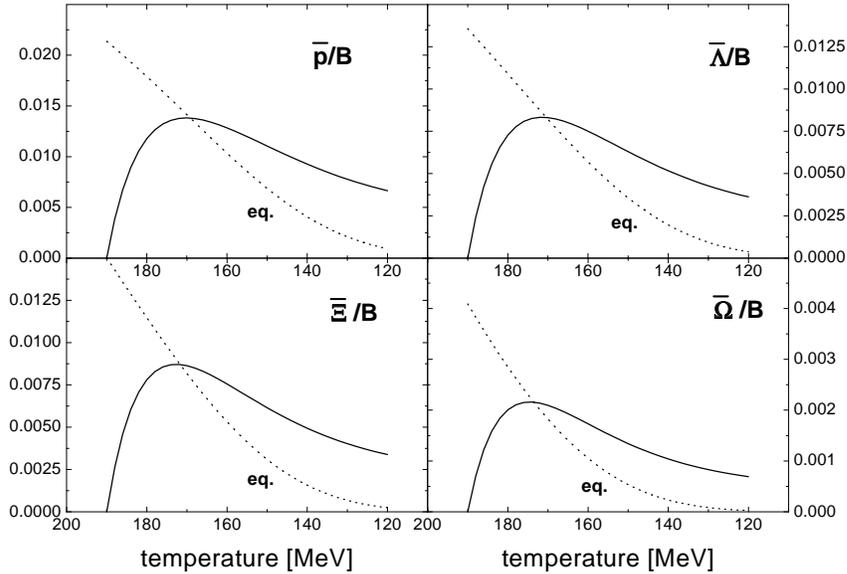}
~\vskip -9mm
\centering
  \caption{
  The antihyperon to baryon number ratio
$N_{\bar{Y}}/N_B \,  (T)$ and $N_{\bar{Y}}^{eq.}/N_B (T) \,  $
(dotted line) as a function of the decreasing temperature.
Parameters are the same as in Fig. 6.
}
\label{fig_all}
\end{figure}

In Fig.~\ref{fig_all} the number of antihyperons of each specie
and the direct anti-protons (not coming from any anti-baryon resonance)
are now shown as a function of the decreasing temperature $T(t)$
of the hadronic system.
For comparison the instantaneous equilibrium abundancy
$N_{\bar{Y}}^{eq.}(T(t),\mu_B(t),\mu_s(t))/N_B$ is also given.
As noted, after a fast initial population,
the individual yields of the antihyperons do overshoot
their respective equilibrium number
and then do finally saturate at some slightly smaller value.
Moreover, one notices that
the yields effectively do saturate at a number
which can be compared to an equivalent equilibrium number
at a temperature parameter around $T_{eff} \approx 150-160$ MeV,
being close
to the ones obtained within the various thermal analyses \cite{BMS96}.
To be more quantitative in comparison with data,
one has at midrapidity the WA97 results
$\bar{\Xi }^+/\bar{\Lambda } \equiv
\bar{\Xi }^+/(\bar{\Lambda }+ \bar{\Sigma }^0) \approx 0.188 \pm 0.016$
and
$\bar{\Omega }^+/\bar{\Xi }^+ = 0.281 \pm 0.053$
\cite{WA97}. Assuming that $N_{\bar{\Lambda } } \approx N_{\bar{\Sigma }^0}$
and $N_{\bar{\Xi }^+ } \approx N_{\bar{\Xi }^0}$
we have (from Fig.~\ref{fig_all})
$\, \bar{\Xi }^+/(\bar{\Lambda }+ \bar{\Sigma }^0) \rightarrow
1/4 *\bar{\Xi }/\bar{\Lambda } \approx 0.22 $
and
$\bar{\Omega }^+/\bar{\Xi }^+ \rightarrow
2 * \bar{\Omega }/\bar{\Xi }
\approx 0.35 $, both values being in good agreement to the data.

\begin{figure}[t]
\centering
\includegraphics[width=105mm]{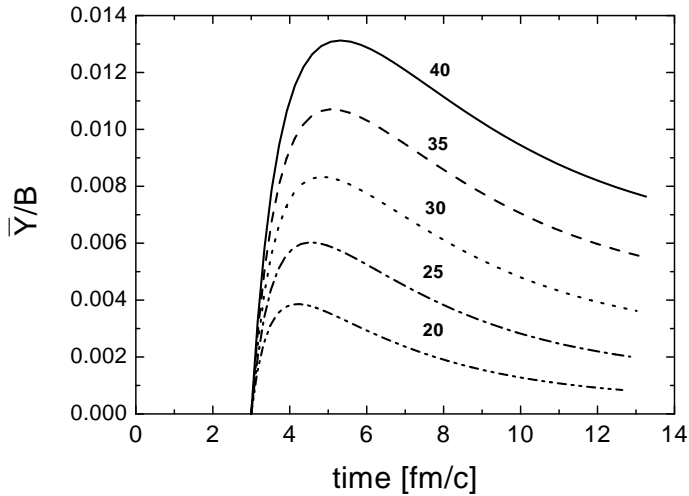}
~\vskip -9mm
\centering
\label{fig_sa}
\caption{$N_{\bar{\Lambda }}/N_B \,(t) $ as a function of time
     for various entropy content described
  via the entropy per baryon ratio ($S/A=20-40$).
  Other parameters are as in Fig.~6.}
\label{fig_entr}
\end{figure}

In Fig.~\ref{fig_entr} the number of anti-$\Lambda $s
as a function of time is given for various entropy per baryon ratios.
One notices that the final value in the yield significantly
depends on the entropy content, or, in other
words, on the baryochemical potential. We note that the absolute
yield on $\bar{\Lambda }$s
at midrapidity from WA97 and NA49 can best be reproduced by employing
an entropy to baryon ratio $S/A\approx 40$:
From WA97 we have about $1.8 \pm 0.2 $\cite{BMS96} and from NA49 about
1.5 \cite{Mischke} anti-Lambdas (and $\bar{\Sigma }^0$s) in
one unit of rapidity.
Assuming again that $N_{\bar{\Lambda } } \approx N_{\bar{\Sigma }^0}$
and taking for the net baryon rapidity distribution
$dN_B /dy \approx 80 $ one has $N_{\bar{\Lambda }}/N_B \approx (1.5/2)/80
\approx 0.0094$, which
would be roughly the outcome for $S/A$
being slightly larger than 40 (compare Fig.~\ref{fig_all}).
Indeed, at
midrapidity one qualitatively
expects a higher entropy content due to the
larger pion to baryon ratio as compared to
full `$4\pi$' data over all rapidities. At this point it will also
be very interesting to compare our semi-quantitative
calculations with the new results from NA49
on the $\bar{\Lambda }$-yield at lower SPS energies of 80
AGeV and 40 AGeV with lower entropy contents, respectively \cite{Mischke}.

To summarize,
multi-mesonic production of antihyperons is a consequence
of detailed balance and, as the annihilation rate is
large, it is by far the most dominant source
in a hadronic gas.
It clearly demonstrates
the importance of hadronic multi-particle channels,
occuring frequently enough in a (moderately) dense hadronic
environment in order to populate and chemically saturate
the rare antibaryons.
In order to be more competitive for a direct
comparison with detailed experimental findings
(like $A_{part}$ dependence),
new strategies have to be developed to describe for such
multi-particle interactions within present day transport codes.
A significant first step forward was very recently
been pursued by Cassing, where the concept of rate calculations
in subdivided, rather local space-time cells has been adopted.
The annihilation cross section is modelled by
three-body vector meson decay.
First results concerning the production of anti-protons
at AGS and SPS energies are quite impressive \cite{Cassing}.
(At SPS energies the $\bar{p}$-yield was calculated to be
nearly a factor of two below the published, yet rather
old data. A new experimental analysis, but
still not presented in public, shows that
the actual experimental $\bar{p}$-yield will be reduced by nearly
a factor of two \cite{Seyboth}. This would mark the
importance of this specific calculation.)
Another strategy for implementing
the backreaction could be to incorporate microscopically
the concept of two meson doorway states \cite{JV88}
and their sequential decay
by standard binary scattering processes.
\\[5mm]
{\bf Acknowledgments}
\\[3mm]
The study presented in section \ref{sec2} was done together
with Bj\"orn Schenke. The author is also indebted
to E.~Bratkovskaya and H.~Weber for illuminating discussions.
This work was supported by
the Bundesministerium f\"ur Bildung und Forschung,
by the Deutsche Forschungs Gemeinschaft and by the
Gesellschaft f\"ur Schwerionenforschung.


\begin{thebibliography}{99}

\bibitem{Bravina} M.~Belkacem et al, \Journal{PRC}{58}{1727}{1998};
L.~Bravina et al, \Journal{\PLB}{434}{379}{1998};
\Journal{PRC}{60}{024904}{1999};
J.~Sollfrank, U.~Heinz, H.~Sorge and N.~Xu,
\Journal{PRC}{59}{1637}{1999}.

\bibitem{BCG00} E.~Bratkovskaya, W.~Cassing, C.~Greiner, M.~Effenberger,
U.~Mosel and A.~Sibirtsev, \Journal{\NPA}{674}{249}{2000}.

\bibitem{BMS96} P.~Braun-Munzinger, I.~Heppe and J.~Stachel,
\Journal{\PLB}{465}{1}{1999}; F.~Becattini, J.~Cleymans,
A.~Ker\"anen, E.~Suhonen and K.~Redlich, \Journal{PRC}{64}{024901}{2001}.

\bibitem{GL00} C.~Greiner and S.~Leupold, \Journal{\JP}{27}{L95}{2001};
C.~Greiner, \Journal{\NPA}{698}{591}{2002}; and
{\em arXiv:nucl-th/0208080 }.

\bibitem{SG02} B.~Schenke and C.~Greiner, in preparation.

\bibitem{Neise} L.~Neise, GSI report 90-24 (1990), unpublished.

\bibitem{So98} H.~Sorge, \Journal{\NPA}{630}{522c}{1998}.

\bibitem{Bass} S.A.~Bass, \Journal{\JP}{28}{1543}{2002}.

\bibitem{CG01} C.~Greiner, \Journal{\JP}{28}{1631}{2002}.

\bibitem{Ge98} J.~Geiss, W.~Cassing and C.~Greiner,
\Journal{\NPA}{644}{107}{1998}.

\bibitem{Weber} H.~Weber, E.~Bratkovskaya and H.~St\"ocker,
{\em arXiv:nucl-th/0205030}; H.~Weber, E.~Bratkovskaya
et al, in preparation.

\bibitem{So95} H.~Sorge et al, \Journal{\PLB}{289}{6}{1992};
H.~Sorge, \Journal{\ZPC}{67}{479}{1995};
\Journal{\PRC}{52}{3291}{1995};
K.~Werner and J.~Aichelin,
\Journal{\PLB}{300}{158}{1993} and \Journal{\PLB}{308}{372}{1993};
N.~Armesto, M.A.~Braun, E.G.~Ferreiro and C.~Pajares,
\Journal{\PLB}{344}{301}{1995};
E.G.~Ferreiro and C.~Pajares,
\Journal{\ZPC}{73}{309}{1997};
M.~Bleicher et al, \Journal{\PLB}{485}{133}{2000}.

\bibitem{RS00} R.~Rapp and E.~Shuryak, \Journal{\PRL}{86}{2980}{2001}.

\bibitem{WA97} R.~Caliandro for the WA97 collabortion,
\Journal{\JP}{25}{171}{1999}.

\bibitem{Mischke} A.~Mischke, these contributions.

\bibitem{Cassing} W.~Cassing, \Journal{\NPA}{700}{618}{2002}.

\bibitem{Seyboth} P.~Seyboth, private communications.

\bibitem{JV88} J.~Vandermeulen, \Journal{ZPC}{37}{563}{1988}.

\end{thebibliography}
\end{document}